\documentclass[twoside, epsfig]{article}

\input oejv.sty

\usepackage[T1]{fontenc}
\usepackage{caption}

\def\arraystretch{1.1}
\def\f''{\hbox{$.\!\!''$}}

\begin{document}


\OEJVhead{September 2017}
\OEJVtitle{CzeV -- The Czech Variable Star catalogue}
\OEJVauth{Skarka, M.$^{1,2}$, Ma\v{s}ek, M.$^{2,3}$, Br\'{a}t, L.$^{2,4}$, Caga\v{s}, Pa.$^{2,5}$, Jury\v{s}ek, J.$^{2,3,6}$, Ho\v{n}kov\'{a}, K.$^{2}$,}\vspace{-0.45cm}
\OEJVauth{Zejda, M.$^{2,7}$, \v{S}melcer, L.$^{2,41}$, Jel\'{i}nek, M.$^{8}$, Lomoz, F.$^{2,9}$, Tyl\v{s}ar, M.$^{15}$, Trnka, J.$^{2,10}$,}\vspace{-0.45cm}
\OEJVauth{Pejcha, O.$^{2,11}$, Pintr, P.$^{2,12}$, Lehk\'{y}, M.$^{2,13}$, Jan\'{i}k, J.$^{7}$, \v{C}ervinka, L.$^{14}$, P\v{r}ib\'{i}k, V.$^{2,16}$,}\vspace{-0.45cm}
\OEJVauth{Motl, D.$^{17}$, Walter, F.$^{2,18}$, Zasche, P.$^{6}$, Koss, K.$^{38}$, H\'{a}jek, P.$^{19}$, B\'{i}lek, F.$^{2,20}$, Li\v{s}ka, J.$^{2,21}$,}\vspace{-0.45cm}
\OEJVauth{Ku\v{c}\'{a}kov\'{a}, H.$^{2,6,8,22,26}$, Bodn\'{a}r, F.$^{23}$, Ber\'{a}nek, J.$^{23}$, \v{S}af\'{a}\v{r}, J.$^{17}$, Moudr\'{a}, M.$^{18}$, Or\v{s}ul\'{a}k, M.$^{23}$,}\vspace{-0.45cm}
\OEJVauth{Pintr, M.$^{12}$, Sobotka, P.$^{2}$, D\v{r}ev\v{e}n\'{y}, R.$^{2,24}$, Jur\'{a}\v{n}ov\'{a}, A.$^{7}$, Pol\'{a}k, J.$^{25}$, Polster, J.$^{7}$, }\vspace{-0.45cm}
\OEJVauth{Onderkov\'{a}, K.$^{26}$, Smolka, M.$^{2,27}$, Auer, R. F.$^{2,28}$, Koci\'{a}n, R.$^{2,26}$, Hlad\'{i}k, B.$^{2,29}$, Caga\v{s}, Pe.$^{30}$,}\vspace{-0.45cm}
\OEJVauth{Gre\v{s}, A.$^{31}$, M\"uller, D.$^{32}$, \v{C}{a}pkov\'{a}, H.$^{13}$, Kysel\'{y}, J.$^{33}$, Hornoch, K.$^{8}$, Truparov\'{a}, K.$^{26}$,}\vspace{-0.45cm}
\OEJVauth{Timko, L.$^{34}$, Bro\v{z}, M.$^{6}$, B\'{i}lek, M.$^{6,8}$, \v{S}ebela, P.$^{35}$, Han\v{z}l, D.$^{36}$, \v{Z}ampachov\'{a}, E.$^{7}$,}\vspace{-0.45cm}
\OEJVauth{Seck\'{a}, J.$^{13}$, Pravec, P.$^{8}$, Mr\v{n}\'{a}k, P.$^{13}$, Svoboda, P.$^{37}$, Ehrenberger, R.$^{2}$, Novotn\'{y}, F.$^{8,39}$,}\vspace{-0.45cm}
\OEJVauth{Poddan\'{y}, S.$^{2,18}$, Prudil, Z.$^{2,40}$, Kuch\v{t}\'{a}k, B.$^{2}$, \v{S}tegner, D.$^{2,7}$}

\OEJVinst{Konkoly Observatory, MTA CSFK, Konkoly Thege M. u. 15--17, H-1121 Budapest, Hungary; {\tt \href{mailto:marek.skarka@mta.csfk.hu}{marek.skarka@mta.csfk.hu}}}
\OEJVinst{Variable Star and Exoplanet Section of the Czech Astronomical Society, Vset\'{i}nsk\'{a} 941/78,\\ CZ-757 01 Vala\v{s}sk\'{e} Mezi\v{r}\'{i}\v{c}\'{i}, Czech Republic}
\OEJVinst{Institute of Physics Czech Academy of Sciences, Na Slovance 1999/2, CZ-182 21 Praha, Czech Republic}
\OEJVinst{ALTAN.Observatory, Velk\'{a} \'{U}pa 193, CZ-542 21 Pec pod Sne\v{z}kou, Czech Republic}
\OEJVinst{BSObservatory, Modr\'{a} 587, CZ-760 01 Zl\'{i}n, Czech Republic}
\OEJVinst{Astronomical Institute, Faculty of Mathematics and Physics, Charles University in Prague, \\V Hole\v{s}ovi\v{c}k\'{a}ch 2, CZ-180 00 Praha 8, Czech Republic}
\OEJVinst{Department of Theoretical Physics and Astrophysics, Masaryk University, Kotl\'{a}\v{r}sk\'{a} 2, CZ-611 37 Brno,\\ Czech Republic}
\OEJVinst{Astronomical Institute, The Czech Academy of Sciences, Fri\v{c}ova 298, CZ-251 65 Ond\v{r}ejov, Czech Republic}
\OEJVinst{Private Observatory, \v{S}vermova 441, CZ-264 01 Sedl\v{c}any, Czech Republic}
\OEJVinst{City Observatory Slan\'{y}, Nosa\v{c}ick\'{a} 1713, CZ-274 01 Slan\'{y}, Czech Republic}
\OEJVinst{Department of Astrophysical Sciences, Princeton University, 4 Ivy Lane, Princeton, NJ 08540, USA}
\OEJVinst{Private Observatory, Svatov\'{a}clavsk\'{a} 2517, CZ-438 01 \v{Z}atec, Czech Republic}
\OEJVinst{\'{U}pice Observatory, U Lipek 160, CZ-542 32 \'{U}pice, Czech Republic}
\OEJVinst{Private Observatory, Svojs\'{i}kova 1370, CZ-293 01 Mlad\'{a} Boleslav, Czech Republic}
\OEJVinst{Prost\v{e}jov Observatory, Riegrova 3348, CZ-796 01 Prost\v{e}jov, Czech Republic}
\OEJVinst{Hinata Observatory, T\v{r}. 3. kv\v{e}tna 689, CZ-763 02 Zl\'{i}n, Czech Republic}
\OEJVinst{Observatory and planetarium Brno, Krav\'{i} hora 2, CZ-616 00 Brno, Czech Republic}
\OEJVinst{\v{S}tef\'{a}nik Observatory, Pet\v{r}\'{i}n 205, CZ-118 46 Praha 1, Czech Republic}
\OEJVinst{MontePa, Pavlovice u Vy\v{s}kova, CZ-683 41 Pavlovice u Vy\v{s}kova, Czech Republic}
\OEJVinst{TS Observatory, Trocnovsk\'{a} 1188, CZ-374 01 Trhov\'{e} Sviny, Czech Republic}
\OEJVinst{Central European Institute of Technology - Brno University of Technology (CEITEC BUT), Purky\v{n}ova
656/123, CZ-612 00 Brno, Czech Republic}
\OEJVinst{Institute of Physics, Faculty of Philosophy and Science, Silesian University in Opava, Bezru\v{c}ovo n\'{a}m. 13, CZ-746 01 Opava, Czech Republic}
\OEJVinst{Gymn\'{a}zium 5. kv\v{e}tna 620, p. o., CZ-432 01 Kada\v{n}, Czech Republic}
\OEJVinst{Znojmo Observatory, Vinohrady 57, CZ-669 02 Znojmo, Czech Republic}
\OEJVinst{Plze\v{n} Observatory, U Dr\'{a}hy 11, CZ-318 00 Plze\v{n}, Czech Republic}
\OEJVinst{Ostrava Planetarium, VSB - Technical University Ostrava, K Planet\'{a}riu 502, CZ-725 26 Ostrava, Czech Republic }
\OEJVinst{Private Observatory, Stani\v{c}n\'{a} 597, Tren\v{c}ianska Turn\'{a} SK-913 21, Slovak Republic}
\OEJVinst{South-Moravian Observatory, Chud\v{c}ice 273, CZ-664 71 Veversk\'{a} Bit\'{y}\v{s}ka, Czech Republic}
\OEJVinst{Private Observatory, Bore\v{c}kova 1422, CZ-198 00 Praha 9, Czech Republic}
\OEJVinst{Kevin T. Crofton Department of Aerospace and Ocean Engineering, Virginia Tech, Randolph Hall, 460 Old Turner St., Blacksburg, VA 24061, USA}
\OEJVinst{Department of Physics Education, Faculty of Mathematics and Physics, Charles University, V Hole\v{s}ovick\'{a}ch 2, 180 00 Praha}
\OEJVinst{Gymn\'{a}zium Bud\v{e}jovick\'{a} 680/17, CZ-140 00 Praha 4, Czech Republic }
\OEJVinst{Institute of Atmospheric Physics AS \v{C}R, Bo\v{c}n\'{i} II 1401, CZ-141 31 Praha, Czech Republic}
\OEJVinst{Franti\v{s}ek Pe\v{s}ta Observatory, Ke Hv\v{e}zd\'{a}rn\v{e} 668, CZ-391 02 Sezimovo \'{U}st\'{i}, Czech Republic}
\OEJVinst{Private Observatory, Senet\'{a}\v{r}ov 206, CZ-679 06 Jedovnice, Czech Republic}
\OEJVinst{Private Observatory, \'{U}voz 118, CZ-602 00 Brno, Czech Republic}
\OEJVinst{Private Observatory, V\'{y}pustky 5, CZ-614 00 Brno, Czech Republic}
\OEJVinst{Observatory Vy\v{s}kov, Krom\v{e}\v{r}\'{i}\v{z}sk\'{a} 721, CZ-682 01 Vy\v{s}kov, Czech Republic}
\OEJVinst{Gymn\'{a}zium Jihlava, J. Masaryka 1560/1, CZ-586 01 Jihlava, Czech Republic}
\OEJVinst{Astronomisches Rechen-Institut, Zentrum f\H{u}r Astronomie der Universit\"{a}t Heidelberg, M\"{o}nchhofstr. 12-14, D-691 20 Heidelberg, Germany}
\OEJVinst{Observatory Vala\v{s}sk\'{e} Mezi\v{r}\'{i}\v{c}\'{i}, Vset\'{i}nsk\'{a} 941/78, CZ-757 01 Vala\v{s}sk\'{e} Mezi\v{r}\'{i}\v{c}\'{i}, Czech~Republic}


\OEJVabstract{We present the first release of the Czech Variable star catalogue that currently contains 1228 stars whose variability was discovered by 60 Czech observers. The catalogue contains confirmed variable stars of various types, but also candidates. We give precise coordinates, cross identification with other catalogues, information about constellation, variability type, brightness, light elements, name of the discoverer and year of discovery. In eighty-eight percent of stars the variability type is estimated, for more than 60\,\% of the stars the light ephemerides are given.}



\begintext
\section{Introduction}\label{Sec:Introduction}
\begin{figure}[htbp]
\centering
\includegraphics[width=0.95\textwidth]{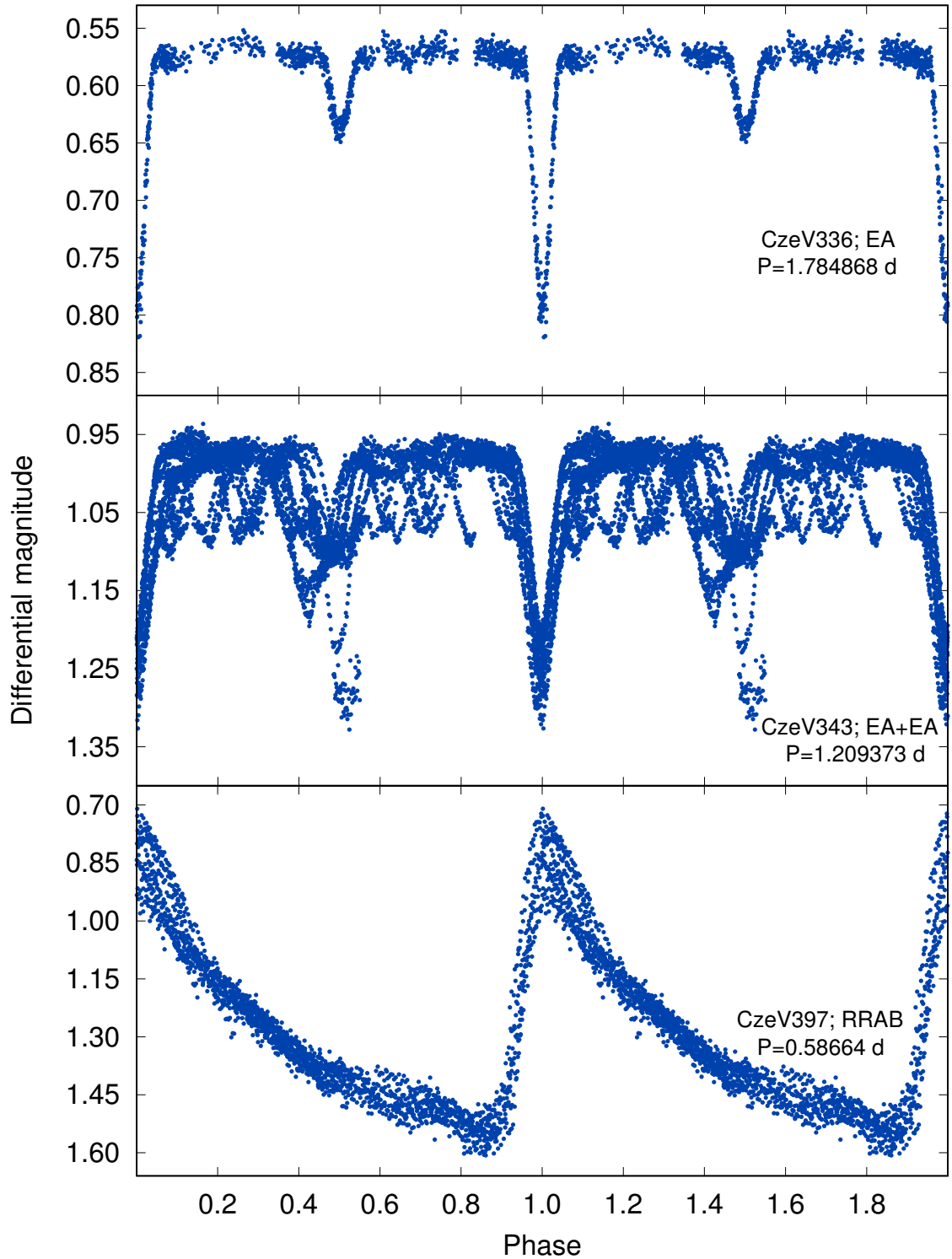}\\
\caption{Examples of three light curves of CzeV stars. The top panel shows an Algol type eclipsing binary CzeV336 \citep{jurysek2012}, the middle panel shows CzeV343, which is a double EA system \citep{cagas2012}, and the bottom panel shows an RR Lyrae type star CzeV397, which undergoes distinct amplitude modulation \citep{skarka2013}.}
\label{Fig:Examples}
\end{figure}

The Czech Variable star catalogue (CzeV) was created by L. \citet{brat2005,brat2006} as a public online database that serves as a list of variable stars whose variability was discovered by observers from the Czech Republic. The initial idea comes from M. Zejda, who compiled Czech discoveries till the establishment of the CzeV catalogue. The CzeV is managed and administrated by the Variable Star and Exoplanet Section of the Czech Astronomical Society (VSES CAS) and is available at \url{http://var2.astro.cz/czev.php}. Every observer registered at the VSES web page\footnote{ \url{http://var2.astro.cz}} can include his/her new variable star to the catalogue. The necessary conditions are that the star shows apparent change in brightness, at the time of discovery was not listed as variable star in the General Catalogue of Variable stars \citep{samus2017} or Variable Star IndeX \citep{watson2006}, and has unambiguous coordinates and designation from photometric/astrometric catalogues. No confirmation by other observer is required. Newly submitted items are checked by the administrator. Because overall publication focused on the CzeV catalogue has been published only now, many of the CzeV stars have been re-discovered by other authors, discoveries already published and stars included to the VSX. Three examples of light curves of CzeV stars are shown in Fig. \ref{Fig:Examples}.

Most stars in the catalogue have been discovered during observation of other variable stars by happy coincidence, not as the main purpose of the observation or organized campaign. This is also the case of discoveries made by the FRAM telescope that is primarily dedicated for extinction measurements \citep{fram}. However, discoveries made by Pavel Caga\v{s} come from a project that is intentionally dedicated for search for new variable stars \citep[e.g. regular long-term observation of the field around V729 Aql,][]{cagas2017}.  

The current version of the CzeV presented here is revised and can slightly differ in details from the online version which will be updated accordingly soon\footnote{ Later improvement of elements, variability types or adding other parameters by registered users of VSES is also possible.}. This discrepancy is given by the history of the catalogue, which was originally intended as an internal list that should unify the preliminary personal identifications (IDs) under a common ID `CzeV'. The entries have not been checked and corrected regularly till 2014. Together with increasing number of newly discovered variable stars, the announcement to the world-wide community become desirable. Because the entries were included by different observers often in wrong format\footnote{ The default and correct format is CzeVnumber, where `number' is the ordinal number in the catalogue.} and entries could contain inaccuracies, the whole catalogue was carefully revised item by item. However, after proper identification of the stars\footnote{ We used Aladin sky atlas \citep{bonnarel2000}.} the automated procedures were used where it was possible to reduce the errors caused by a human factor (e.g. coordinates and magnitudes extraction from the catalogues). The whole catalogue is now unified in its form and all entries are in the same format. 

\section{Description of the catalogue}\label{Sect:Description}
For a smooth transfer to the VSX we followed their requirements on the format and other necessary information. The whole table containing all entries is placed at the end of the paper (Table \ref{Tab:FullTable}) and is attached in a machine-readable format (.csv file) as a supplementary material to this paper. The table contains the following columns:
\begin{enumerate}
	\item {\it CzeV number; `CzeV'} -- The ordinal number in the catalogue. Because there has not been any input control for the presence of the objects in CzeV, there are 17 items that are doubled. In two cases the identification of stars was ambiguous and therefore the corresponding rows are left blank. The items listed in the first column are linked with the online version and take the reader to the web page where the light curve and chart can be examined. 
	\item {\it Cross identification; `ID'} -- we give UCAC4 \citep{zacharias2013} as the default cross-ID (1048 entries). If no UCAC4 ID is available (156 objects), we give USNO-B1.0 ID. In five cases none of these IDs is available. Thus, we give 2MASS ID \citep{cutri2003}. 
	\item {\it VSX identification; `VSX'} -- If the star is known to the VSX, then we give `1'\footnote{ These items are linked with the online VSX catalogue.} in this column (261 stars including 62 with CzeV~ID). In the machine-readable file the full VSX designation is given in this column.
	\item {\it Right ascension; `RA'} -- Right ascension in [hh mm ss.sss] format, J2000 equinox. The coordinate is taken from the UCAC4. If the star has not UCAC4 ID, the coordinate is taken from the PPMXL catalogue \citep{roeser2010}.
	\item {\it Declination; `DE'} -- Declination in [$\pm$dd mm ss.ss] format, J2000 equinox. The coordinate is taken from the UCAC4. If the star has not UCAC4 ID, the coordinate is taken from the PPMXL catalogue \citep{roeser2010}.
	\item {\it Constellation; `Con'} -- Three-letter standard abbreviation of the constellation.
	\item {\it Variability type; `Type'} -- The type according to VSX nomenclature\footnote{ \url{https://www.aavso.org/vsx/index.php?view=about.vartypes}}. In case the type is only suggested, the type is followed by double colon.  
	\item {\it APASS V magnitude; `V'} -- The mean Johnson $V$ magnitude taken from the UCAC4. The $V$ magnitude can be calculated using various transformations from other photometric systems when APASS $V$ magnitude is unavailable. Such procedure is recommended by the VSX administrators. However, after several tests we realized that various calibrations give results that could differ by several tenths of magnitude. For blue and red stars the calibrations usually give completely absurd results. Thus, we do not give $V$ magnitude when it is not available in UCAC4. 
	\item {\it 2MASS $J$ magnitude; `J'} -- We give 2MASS $J$ magnitude for almost all targets to give at least some information about the brightness of the star when $V$ magnitude is unavailable\footnote{ When also $J$ mag is unavailable, we give USNO-B1.0 $R1$ magnitude here (2 stars).}. 
	\item {\it 2MASS $J-K$; `J-K'} -- The colour index $J-K$ in magnitudes is given for a rough impression about the spectral type of the CzeV stars.
	\item {\it Amplitude; `A'} -- The amplitude of the light variations in magnitudes detected in the passband given in the next column.
	\item {\it Filter; `F'} -- The filter in which the amplitude was estimated.
	\item {\it Zero epoch; `M0'} -- Zero epoch in Julian date (time of minimum for eclipsing binaries and ellipsoidal variables, maximum for pulsating variable stars). Unfortunately, we do not have information about the type of the date (Geocentric vs. Heliocentric). Thus these values should be handled with caution.
	\item {\it Period; `P'} -- Period of the variations in days. The errors are usually unknown and, therefore, not given\footnote{ VSX and GCVS also does not provide errors. Therefore, the absence of errors should not be a problem for inclusion of our targets to these catalogues.}. In spurious cases we give the value with 6 decimal places. \vspace*{-0.1cm}
	\item {\it Discoverer} -- The abbreviation of the discoverer(s). The list is given in Table \ref{Table:Observers}.\vspace*{-0.1cm}
	\item {\it Year of the discovery; `Year'.}
\end{enumerate}   

\section{Statistics on the sample and further details}\label{Sect:Statistics}

\begin{figure}[htbp]
\centering
\includegraphics[width=1.0\textwidth]{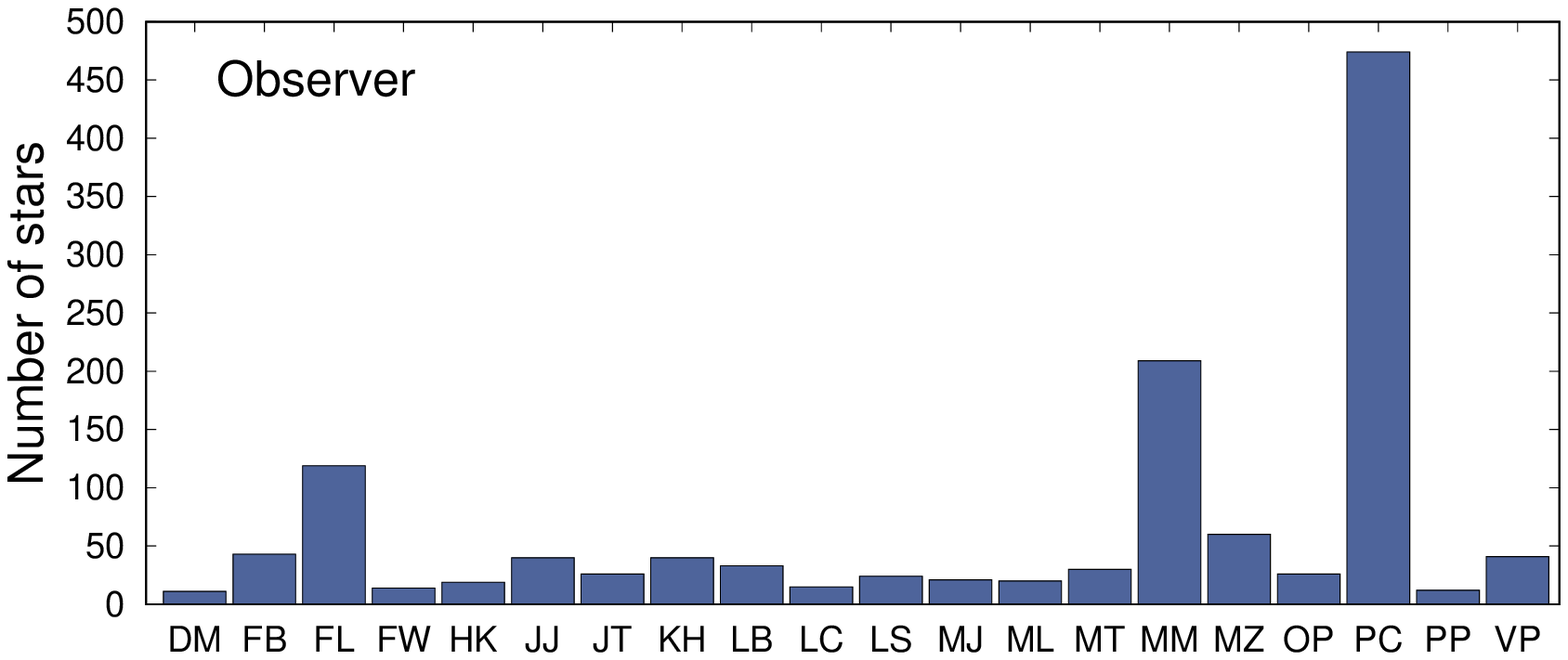}\\
\includegraphics[width=1.0\textwidth]{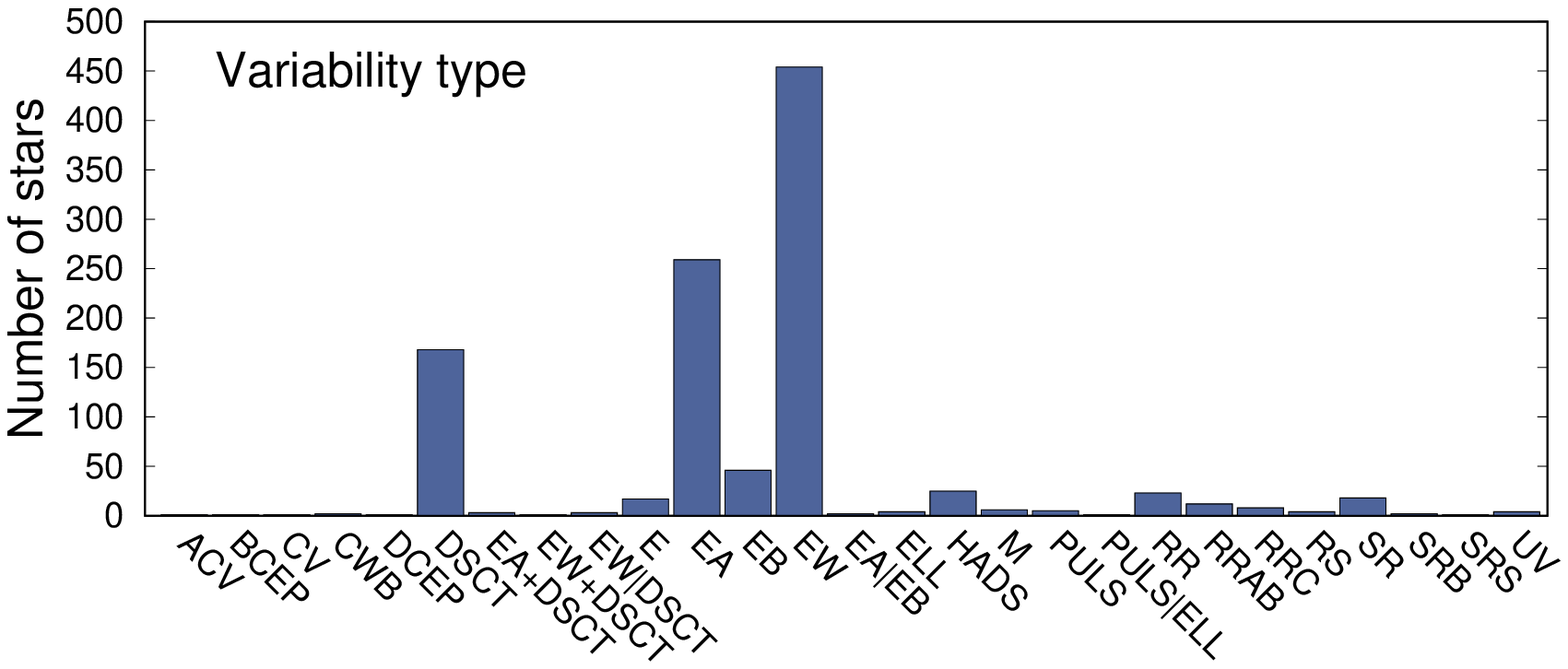}\\
\includegraphics[width=1.0\textwidth]{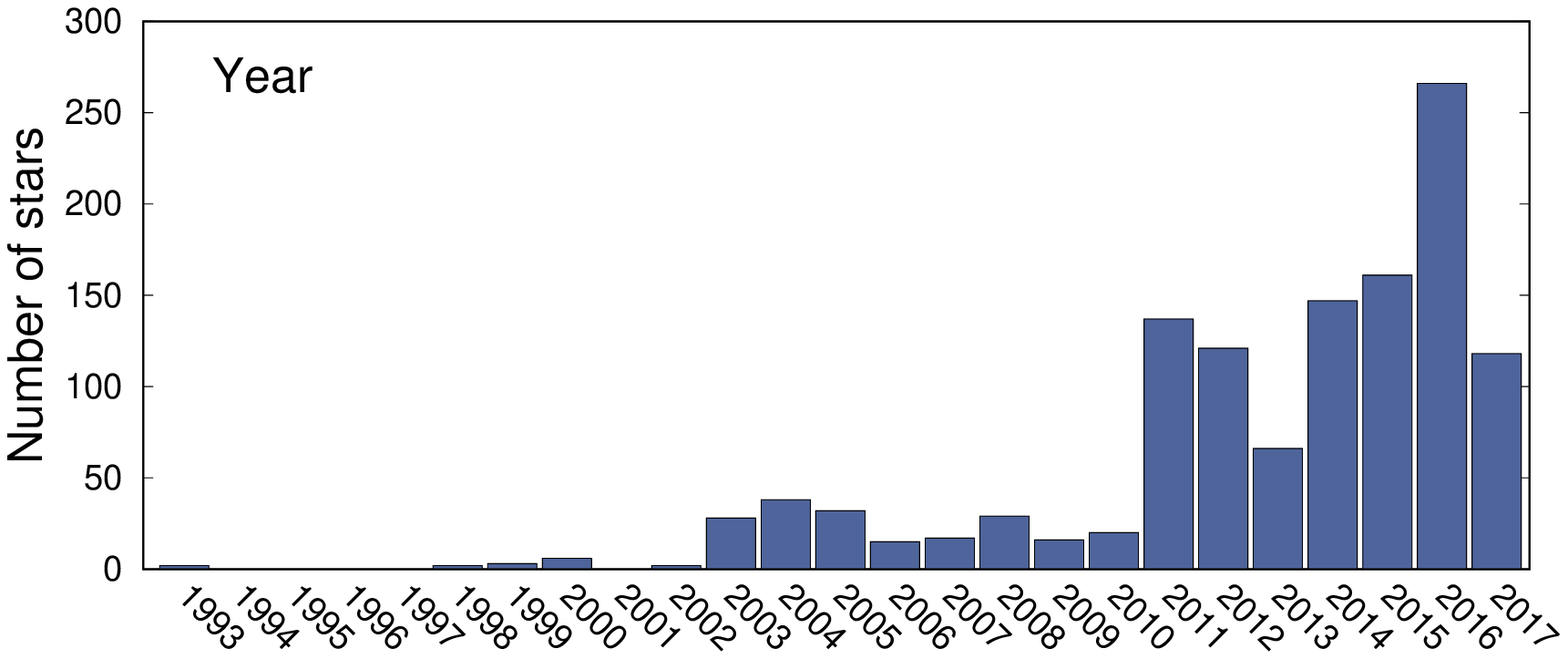}
\caption{The distribution of the number of stars discovered by twenty most productive observers (top panel, the short cuts can be found in Table \ref{Table:Observers}). The middle panel shows numbers of different variability type stars in the CzeV catalogue. Finally, the distribution of the discoveries during time is in the bottom panel.}
\label{Fig:Histograms}
\end{figure}

At the end of June 2017 the catalogue has contained 1228 stars discovered by 60 observers. The most productive observers are Pavel Caga\v{s}, Martin Ma\v{s}ek and Franti\v{s}ek Lomoz (472, 209 and 119 entries, respectively). The complete list of observers is in Table \ref{Table:Observers} and the distribution of discoveries among twenty most productive observers is shown in the top panel of Fig. \ref{Fig:Histograms}. The sum of the numbers in Table \ref{Table:Observers} does not correspond to the total number of discovered variable stars because discoverers often observe in teams. Data are available upon request.

\begin{table}[htbp]
\centering
\caption{Discoverers, their short cuts (Abb.), number of their discoveries (Nr.) and emails.}
\begin{tiny}
\begin{tabular}{llrlllrl}
\hline\hline
Name & Abb. & Nr. & Email & Name & Abb. & Nr. & Email\\ \hline
Reinhold	F.	Auer		&	RFA	&	2	&	\href{mailto:	auer.reinhold@gmail.com	}{	auer.reinhold@gmail.com	}	&	Martin	Ma\v{s}ek	&	MM	&	209	&	\href{mailto:	cassi@astronomie.cz	}{	cassi@astronomie.cz	}	\\
Franti\v{s}ek	B\'{i}lek			&	FB	&	43	&	\href{mailto:	frantabilek@gmail.com	}{	frantabilek@gmail.com	}	&	David	Motl	&	DM	&	11	&	\href{mailto:	dmotl@volny.cz	}{	dmotl@volny.cz	}	\\
Michal	B\'{i}lek			&	MiB	&	1	&	\href{mailto:	michal.bilek@asu.cas.cz	}{	michal.bilek@asu.cas.cz	}	&	Milada	Moudr\'{a}	&	MiM	&	3	&	\href{mailto:	moudra@fzu.cz	}{	moudra@fzu.cz	}	\\
Jan	Ber\'{a}nek			&	JB	&	3	&	\href{mailto:		}{		}	&	Petr	Mr\v{n}\'{a}k	&	PM	&	1	&	\href{mailto:	mrnak.petr@email.cz	}{	mrnak.petr@email.cz	}	\\
Fabi\'{a}n	Bodn\'{a}r			&	FaB	&	3	&	\href{mailto:	fabian.bodnar@seznam.cz	}{	fabian.bodnar@seznam.cz	}	&	Denis	M\H{u}ller	&	DeM	&	1	&	\href{mailto:	topkvark@seznam.cz	}{	topkvark@seznam.cz	}	\\
Lubo\v{s}	Br\'{a}t			&	LB	&	33	&	\href{mailto:	brat@pod.snezkou.cz	}{	brat@pod.snezkou.cz	}	&	Filip	Novotn\'{y}	&	FN	&	1	&	\href{mailto:	fildanovo@gmail.com	}{	fildanovo@gmail.com	}	\\
Miroslav	Bro\v{z}			&	MB	&	1	&	\href{mailto:	mira@sirrah.troja.mff.cuni.cz	}{	mira@sirrah.troja.mff.cuni.cz	}	&	Kate\v{r}ina	Onderkov\'{a}	&	KO	&	2	&	\href{mailto:	katka.onderkova@centrum.cz	}{	katka.onderkova@centrum.cz	}	\\
Pavel	Caga\v{s}			&	PC	&	472	&	\href{mailto:	pavel.cagas@gmail.com	}{	pavel.cagas@gmail.com	}	&	Martin	Or\v{s}ul\'{a}k	&	MO	&	3	&	\href{mailto:	martas.orsulak@gmail.com	}{	martas.orsulak@gmail.com	}	\\
Petr	Caga\v{s}			&	PeC	&	2	&	\href{mailto:	pcagas@vt.edu	}{	pcagas@vt.edu	}	&	V\'{a}clav	P\v{r}ib\'{i}k	&	VP	&	41	&	\href{mailto:	vaclav.pribik@gmail.com	}{	vaclav.pribik@gmail.com	}	\\
Hedvika	\v{C}apkov\'{a}			&	HC	&	3	&	\href{mailto:	hedvika.capkova@gmail.com	}{	hedvika.capkova@gmail.com	}	&	Ond\v{r}ej	Pejcha	&	OP	&	26	&	\href{mailto:	pejcha@astro.princeton.edu	}{	pejcha@astro.princeton.edu	}	\\
Ladislav	\v{C}ervinka			&	LC	&	15	&	\href{mailto:	mail@ladislavcervinka.cz	}{	mail@ladislavcervinka.cz	}	&	Michal	Pintr	&	MP	&	3	&	\href{mailto:	M.Pirati@seznam.cz	}{	M.Pirati@seznam.cz	}	\\
Radek	D\v{r}ev\v{e}n\'{y}			&	RD	&	4	&	\href{mailto:	radek.dreveny@volny.cz	}{	radek.dreveny@volny.cz	}	&	Pavel	Pintr	&	PP	&	12	&	\href{mailto:	pintr@ipp.cas.cz	}{	pintr@ipp.cas.cz	}	\\
Roman	Ehrenberger			&	RE	&	1	&	\href{mailto:	ehrenbergerr@opp.cz	}{	ehrenbergerr@opp.cz	}	&	Ji\v{r}\'{i}	Pol\'{a}k	&	JiP	&	5	&	\href{mailto:	jiri.polak@centrum.cz	}{	jiri.polak@centrum.cz	}	\\
Adam	Gre\v{s}			&	AG	&	1	&	\href{mailto:	adam.gres1@gmail.com	}{	adam.gres1@gmail.com	}	&	Jan	Polster	&	JP	&	4	&	\href{mailto:	jpolster@email.cz	}{	jpolster@email.cz	}	\\
Petr	H\'{a}jek			&	PH	&	9	&	\href{mailto:	hv.hajek@seznam.cz	}{	hv.hajek@seznam.cz	}	&	Petr	Pravec	&	PeP	&	1	&	\href{mailto:	petr.pravec@asu.cas.cz	}{	petr.pravec@asu.cas.cz	}	\\
Dalibor	Han\v{z}l			&	DH	&	5	&	\href{mailto:	hanzl@sci.muni.cz	}{	hanzl@sci.muni.cz	}	&	Jaroslava	Seck\'{a}	&	JaS	&	1	&	\href{mailto:	451559@mail.muni.cz	}{	451559@mail.muni.cz	}	\\
Bohuslav	Hlad\'{i}k			&	BH	&	2	&	\href{mailto:	bohuslav.hladik@email.cz	}{	bohuslav.hladik@email.cz	}	&	Miroslav	Smolka	&	MS	&	2	&	\href{mailto:	miroslav.smolka@gmail.com	}{	miroslav.smolka@gmail.com	}	\\
Kate\v{r}ina	Ho\v{n}kov\'{a}			&	KH	&	40	&	\href{mailto:	katerina.honkova@astronomie.cz	}{	katerina.honkova@astronomie.cz	}	&	Petr	Sobotka	&	PeS	&	2	&	\href{mailto:	sobotka@astro.cz	}{	sobotka@astro.cz	}	\\
Kamil	Hornoch			&	KaH	&	1	&	\href{mailto:	k.hornoch@centrum.cz	}{	k.hornoch@centrum.cz	}	&	Petr	Svoboda	&	PS	&	3	&	\href{mailto:	tribase.net@volny.cz	}{	tribase.net@volny.cz	}	\\
Jan	Jan\'{i}k			&	JaJ	&	8	&	\href{mailto:	honza@physics.muni.cz	}{	honza@physics.muni.cz	}	&	Jan	\v{S}af\'{a}\v{r}	&	JS	&	8	&	\href{mailto:	jan@livephotography.net	}{	jan@livephotography.net	}	\\
Martin	Jel\'{i}nek			&	MJ	&	22	&	\href{mailto:	mates@asu.cas.cz	}{	mates@asu.cas.cz	}	&	Pavel	\v{S}ebela	&	PaS	&	1	&	\href{mailto:	pavel.seb@centrum.cz	}{	pavel.seb@centrum.cz	}	\\
Anna	Jur\'{a}\v{n}ov\'{a}			&	AJ	&	5	&	\href{mailto:	juranova@physics.muni.cz	}{	juranova@physics.muni.cz	}	&	Ladislav	\v{S}melcer	&	LS	&	24	&	\href{mailto:	lsmelcer@astrovm.cz	}{	lsmelcer@astrovm.cz	}	\\
Jakub	Jury\v{s}ek			&	JJ	&	40	&	\href{mailto:	jurysek@fzu.cz	}{	jurysek@fzu.cz	}	&	Luk\'{a}\v{s}	Timko	&	LT	&	1	&	\href{mailto:	timkolukas@seznam.cz	}{	timkolukas@seznam.cz	}	\\
Radek	Koci\'{a}n			&	RK	&	6	&	\href{mailto:	koca@astronomie.cz	}{	koca@astronomie.cz	}	&	Jaroslav	Trnka	&	JT	&	26	&	\href{mailto:	hvezdarna@volny.cz	}{	hvezdarna@volny.cz	}	\\
Karel	Koss			&	KK	&	9	&	\href{mailto:	karel.koss@tiscali.cz	}{	karel.koss@tiscali.cz	}	&	Kamila	Truparov\'{a}	&	KT	&	1	&	\href{mailto:	kamila.truparova@vsb.cz	}{	kamila.truparova@vsb.cz	}	\\
Hana	Ku\v{c}\'{a}kov\'{a}			&	HK	&	19	&	\href{mailto:	Hana.Kucakova@centrum.cz	}{	Hana.Kucakova@centrum.cz	}	&	Martin	Tyl\v{s}ar	&	MT	&	30	&	\href{mailto:	mtylsar@astronomie.cz	}{	mtylsar@astronomie.cz	}	\\
Jan	Kysel\'{y}			&	JK	&	1	&	\href{mailto:	kysely@ufa.cas.cz	}{	kysely@ufa.cas.cz	}	&	Filip	Walter	&	FW	&	14	&	\href{mailto:	edmund.squirrel@seznam.cz	}{	edmund.squirrel@seznam.cz	}	\\
Martin	Lehk\'{y}			&	ML	&	20	&	\href{mailto:	makalaki@astro.sci.muni.cz	}{	makalaki@astro.sci.muni.cz	}	&	Petr	Zasche	&	PZ	&	10	&	\href{mailto:	zasche@sirrah.troja.mff.cuni.cz	}{	zasche@sirrah.troja.mff.cuni.cz	}	\\
Ji\v{r}\'{i}	Li\v{s}ka			&	JL	&	6	&	\href{mailto:	jiriliska@post.cz	}{	jiriliska@post.cz	}	&	Miloslav	Zejda	&	MZ	&	60	&	\href{mailto:	zejda@physics.muni.cz	}{	zejda@physics.muni.cz	}	\\
Franti\v{s}ek	Lomoz			&	FL	&	119	&	\href{mailto:	hvezdarna@tiscali.cz	}{	hvezdarna@tiscali.cz	}	&	Eva	\v{Z}ampachov\'{a}	&	EZ	&	8	&	\href{mailto:	eva.zampachova@seznam.cz	}{	eva.zampachova@seznam.cz	}	\\ \hline

\end{tabular}\label{Table:Observers}
\end{tiny}
\end{table}

The variability type is given for 1074 stars. The numbers of particular variable types are in Table~\ref{Table:VarTypes} and shown in the middle panel of Fig. \ref{Fig:Histograms}. The most numerous classes are W Ursae Majoris stars (EW, 453), Algol-type stars (EA, 260) and $\delta$ Scuti stars (DSCT, 167).

\begin{table}[htbp]
\centering
\caption{Numbers of variable types (Nr.) identified among CzeV stars.}
\begin{tabular}{llllll}
\hline\hline
Type & Nr. & Type & Nr. & Type & Nr. \\ \hline
ACV	&	1	&	E	&	17	&	PULS|ELL	&	1	\\
BCEP	&	1	&	EA	&	260	&	RR	&	23	\\
CV	&	1	&	EB	&	47	&	RRAB	&	13	\\
CWB	&	2	&	EW	&	453	&	RRC	&	8	\\
DCEP	&	1	&	EA|EB	&	2	&	RS	&	4	\\
DSCT	&	167	&	ELL	&	4	&	SR	&	18	\\
EA+DSCT	&	3	&	HADS	&	25	&	SRB	&	2	\\
EW+DSCT	&	1	&	M	&	6	&	SRS	&	1	\\
EW|DSCT	&	3	&	PULS	&	5	&	UV	&	4	\\ \hline

\end{tabular}\label{Table:VarTypes}
\end{table}

\newpage

The first provable discovery of a new variable star by a Czech observer dates back to 1993. Since 2013 the number of new variable stars steadily increases, with maximum in 2016. Till the end of June 2017, 118 new variables has been discovered. The history of the discoveries is apparent from the time distribution in the bottom panel of Fig. \ref{Fig:Histograms} and Table \ref{Table:Years}.

\begin{table}[htbp]
\centering
\caption{Number of stars (Nr.) discovered in given year.}
\begin{tabular}{llllllllll}
\hline\hline
Year	&	Nr.	&	Year	&	Nr.	&	Year	&	Nr.	&	Year	&	Nr.	\\ \hline
1998	&	2	&	2003	&	28	&	2008	&	29	&	2013	&	66	\\
1999	&	3	&	2004	&	38	&	2009	&	16	&	2014	&	147	\\
2000	&	6	&	2005	&	32	&	2010	&	20	&	2015	&	161	\\
2001	&	0	&	2006	&	15	&	2011	&	137	&	2016	&	266	\\
2002	&	2	&	2007	&	17	&	2012	&	121	&	2017	&	118	\\ \hline
\end{tabular}\label{Table:Years}
\end{table}

\section*{Acknowledgements}
MS acknowledges the support of the postdoctoral fellowship programme of the Hungarian Academy of Sciences at the Konkoly Observatory as host institution. The financial support of the Hungarian NKFIH Grants K-115709 is acknowledged. We would like to thank the Pierre Auger Collaboration for the use of its facilities. The operation of the robotic telescope FRAM was supported by the EU grant GLORIA (No. 283783 in FP7-Capacities program) and by
the grants of the Ministry of Education of the Czech Republic (MSMT-CR LM2015038, LTT17006 and LM2015046). The data calibration and analysis related to FRAM telescope is supported by the Ministry of Education of the Czech
Republic MSMT-CR (LG15014, CZ.02.1.01/0.0/0.0/16\_013/0001402) and EU/MSMT CZ.02.1.01/0.0/0.0/16\_013/0001403. ZP acknowledge the support of the Hector Fellow Academy.

\captionsetup{width=1.4\textwidth}
\begin{landscape}

\newcolumntype{L}{>{\scriptsize}l}
\newcolumntype{R}{>{\scriptsize}r}
\newcolumntype{C}{>{\scriptsize}c}
\renewcommand\arraystretch{0.9}
\setlength{\tabcolsep}{2pt}
\begin{longtable}{RLCCCCCCCCCCLLCL}
\caption{The main table. See the text in sect. \ref{Sect:Description} for table description.} \\ \label{Tab:FullTable}
\centering

CzeV	&	\multicolumn{1}{C}{ID}	&	VSX	&	RA	&	DE	&	Con	&	Type	&	$V$	&	$J$	&	$J-K$	&	$A$	&	F	&	\multicolumn{1}{C}{M0}	&	\multicolumn{1}{C}{$P$}	&	\multicolumn{1}{C}{Discoverer}	&	Year	\\  \hline\hline
\endfirsthead
\caption{continued.}\\
\hline\hline

CzeV	&	\multicolumn{1}{C}{ID}	&	VSX	&	RA	&	DE	&	Con	&	Type	&	$V$	&	$J$	&	$J-K$	&	$A$	&	F	&	\multicolumn{1}{C}{M0}	&	\multicolumn{1}{C}{$P$}	&	\multicolumn{1}{C}{Discoverer}	&	Year	\\  \hline\hline
\endhead
\hline
\endfoot

\href{http://var2.astro.cz/czev.php?id=1}{1}	&	UCAC4 537-046686	&	\href{https://www.aavso.org/vsx/index.php?view=detail.top&oid=4974}{1}	&	08 29 39.312	&	+17 17 00.58	&	Cnc	&	EA	&	10.77	&	9.384	&	0.531	&	0.57	&	$	C	$	&	2452500.939	&	1.469331	&	PeP	&	1993	\\
\href{http://var2.astro.cz/czev.php?id=2}{2}	&	UCAC4 797-019460	&	\href{https://www.aavso.org/vsx/index.php?view=detail.top&oid=37266}{1}	&	09 54 28.620	&	+69 13 22.28	&	UMa	&	EW	&	11	&	10.194	&	0.251	&	0.39	&	$	V	$	&	2454829.6982	&	0.528847	&	KaH, JK	&	1993	\\
\href{http://var2.astro.cz/czev.php?id=3}{3}	&	UCAC4 655-102227	&		&	21 36 09.916	&	+40 52 39.07	&	Cyg	&		&	10.7	&	7.978	&	0.936	&		&				&		&		&	MZ	&	1998	\\
\href{http://var2.astro.cz/czev.php?id=4}{4}	&	UCAC4 638-091827	&	\href{https://www.aavso.org/vsx/index.php?view=detail.top&oid=13160}{1}	&	20 15 55.943	&	+37 27 15.53	&	Cyg	&	EW	&	11.94	&	11.185	&	0.236	&		&				&	2456518.9723	&	0.404161	&	JS	&	1999	\\
\href{http://var2.astro.cz/czev.php?id=5}{5}	&	UCAC4 638-091516	&	\href{https://www.aavso.org/vsx/index.php?view=detail.top&oid=13159}{1}	&	20 15 17.569	&	+37 31 43.92	&	Cyg	&	EA	&	11.95	&	10.845	&	0.318	&		&				&	2451427.4063	&	0.610595	&	JS	&	1999	\\
\href{http://var2.astro.cz/czev.php?id=6}{6}	&	UCAC4 450-086568	&		&	18 55 45.481	&	-00 00 44.33	&	Aql	&		&	11.97	&	9.698	&	0.387	&		&				&		&		&	MZ	&	1999	\\
\href{http://var2.astro.cz/czev.php?id=7}{7}	&	UCAC4 629-008536	&	\href{https://www.aavso.org/vsx/index.php?view=detail.top&oid=136050}{1}	&	02 41 41.016	&	+35 42 54.87	&	Per	&	EW	&	13.58	&	12.591	&	0.293	&	0.58	&	$	C	$	&	2452996.6731	&	0.373973	&	MZ	&	2000	\\
\href{http://var2.astro.cz/czev.php?id=8}{8}	&	UCAC4 504-106930	&		&	19 31 30.415	&	+10 47 08.69	&	Aql	&		&	12.32	&	11.278	&	0.356	&		&				&		&		&	MZ	&	2000	\\
\end{longtable}

\end{landscape}

\end{document}